\documentclass[10pt,aps,prl,floatfix,superscriptaddress,nofootinbib,notitlepage,reprint]{revtex4-1}
\usepackage{amsmath,amssymb}
\usepackage{graphicx}
\usepackage[dvipsnames]{xcolor}
\usepackage[colorlinks=true,allcolors=BlueViolet]{hyperref}

\usepackage{multirow}
\usepackage{epstopdf}
\usepackage{lipsum}
\usepackage{tikz}
\usetikzlibrary{positioning}
\usetikzlibrary{calc}
\usepackage{lmodern}
\usepackage[T1]{fontenc}

\newcommand{\cou}[1]{\left\lvert g_{#1} \right\rvert}

\begin{document}
\title{\boldmath Two-pole structure of the $D^\ast_0(2400)$}
\author{Miguel~Albaladejo}
\affiliation{Instituto de F\'isica Corpuscular (IFIC),
             Centro Mixto CSIC-Universidad de Valencia,
             Institutos de Investigaci\'on de Paterna,
             Aptd. 22085, E-46071 Valencia, Spain}

\author{Pedro~Fernandez-Soler}
\affiliation{Instituto de F\'isica Corpuscular (IFIC),
             Centro Mixto CSIC-Universidad de Valencia,
             Institutos de Investigaci\'on de Paterna,
             Aptd. 22085, E-46071 Valencia, Spain}

\author{Feng-Kun~Guo}
\affiliation{CAS Key Laboratory of Theoretical Physics, Institute of Theoretical Physics,
Chinese Academy of Sciences, Beijing 100190, China}

\author{Juan~Nieves}
\affiliation{Instituto de F\'isica Corpuscular (IFIC),
             Centro Mixto CSIC-Universidad de Valencia,
             Institutos de Investigaci\'on de Paterna,
             Aptd. 22085, E-46071 Valencia, Spain}
\date{\today}

\begin{abstract}

The so far only known charmed non-strange scalar meson is dubbed as
$D_0^*(2400)$ in the Review of Particle Physics. We show, within the
framework of unitarized chiral perturbation theory, that there are in
fact two $(I=1/2, J^P=0^+)$ poles in the region of the $D_0^*(2400)$
in the coupled-channel $D\pi$, $D\eta$ and $D_s\bar K$ scattering
amplitudes. With all the parameters previously fixed, we predict the
energy levels for the coupled-channel system in a finite volume, and
find that they agree remarkably well with recent lattice QCD
calculations. This successful description of the lattice data is
regarded as a strong evidence for the two-pole structure of the
$D_0^*(2400)$. With the physical quark masses, the poles are located
at $\left(2105^{+6}_{-8}-i\,102^{+10}_{-12}\right)$~MeV and
$\left(2451^{+36}_{-26}-i\,134^{+7}_{-8}\right)$~MeV, with the largest
couplings to the $D\pi$ and $D_s\bar K$ channels, respectively. Since
the higher pole is close to the $D_s\bar K$ threshold, we expect it to
show up as a threshold enhancement in the $D_s\bar K$ invariant mass
distribution. This could be checked by high-statistic data in future
experiments. We also show that the lower pole belongs to the same
SU(3) multiplet as the $D_{s0}^*(2317)$ state.  Predictions for
partners in the bottom sector are also given.

\end{abstract}

\maketitle

\paragraph{Introduction.---} In our quest for understanding the
fundamental theory of the strong interactions, Quantum Chromodynamics
(QCD), the interpretation of the hadronic spectrum plays a key
role. The latter has received a renewed interest with the recent
advent of powerful experimental facilities, capable of exploring open
and hidden charm (and bottom) energy ranges, as reflected in the
Particle Data Group~(PDG)~\cite{PDG2016}. The $D^\ast_{s0}(2317)$
\cite{Aubert:2003fg} and $D^\ast_0(2400)$ \cite{Abe:2003zm} are the
lightest scalar ($J^P = 0^+$) charm strange and non-strange mesons,
respectively, and have attracted much attention~\cite{Chen:2016spr} as
their masses differ from the quark model expectations for $1P$
states. The $D^\ast_{s0}(2317)$ is around $150\ \text{MeV}$ below the
predicted mass~\cite{Godfrey:1985xj,Godfrey:2015dva}~(see, however,
Refs.~\cite{Lakhina:2006fy,Ortega:2016mms}), and one would naively
expect a significantly larger mass for a $c\bar{s}$ than for a $c\bar{n}$, in
contrast with the observation.  Clearly,
an improved understanding of the nature of these resonances is needed.
Different schemes have described them as mostly $c\bar{q}$
states~\cite{Dai:2003yg,Narison:2003td,Bardeen:2003kt,Lee:2004gt,Wang:2006bs},
as mixture of $c\bar{q}$ with tetraquarks~\cite{Browder:2003fk} or
meson-meson~\cite{vanBeveren:2003kd} components, as purely
tetraquarks~\cite{Cheng:2003kg,Terasaki:2003qa,Chen:2004dy,Maiani:2004vq,
  Bracco:2005kt, Wang:2006uba}, or as heavy--light meson
molecules~\cite{Barnes:2003dj,Szczepaniak:2003vy,Kolomeitsev:2003ac,
  Hofmann:2003je,
  Guo:2006fu,Gamermann:2006nm,Faessler:2007gv,Flynn:2007ki,Albaladejo:2016hae}
motivated by the closeness of the $D^\ast_{s0}(2317)$ to the $DK$
threshold.  

There have also been lattice QCD (LQCD) simulations to understand the
charmed scalar sector. Early studies used only $c\bar{s}$
interpolators for the $D_{s0}^*(2317)$, obtaining masses generally
larger than the physical one~\cite{Bali:2003jv,Dougall:2003hv}. Masses
consistent with the $D_{s0}^*(2317)$ and $D^\ast_0(2400)$ experimental
ones were only obtained after including also meson-meson
interpolators~\cite{Mohler:2012na,Mohler:2013rwa}.  The first LQCD
study of $D\pi$, $D\eta$, and $D_s \bar{K}$ coupled-channel scattering
(for $m_\pi \simeq 391\ \text{MeV}$) was recently
reported~\cite{Moir:2016srx}. Therein, a bound state with a large
coupling to $D\pi$ is found and assigned to the $D_0^*(2400)$.

While the $D_{s0}^*(2317)$ is very narrow and its mass is well 
measured~\cite{PDG2016}, the situation for the broad $D_0^*(2400)$ is 
less clear: the reported mass values for the $D_0^*(2400)^0$ at $B$-factories, 
$(2308\pm36)$~MeV (Belle~\cite{Abe:2003zm}) and $(2297\pm22)$~MeV 
(BaBar~\cite{Aubert:2009wg}), differ from that in $\gamma\,A$ reactions, 
$(2407\pm41)$~MeV (FOCUS~\cite{Link:2003bd}), while the LHCb value for the 
charged partner lies in between~\cite{Aaij:2015kqa}. These analyses use 
Breit--Wigner parameterizations and assume a single scalar particle.

Moreover, a better understanding of the $D^\ast_{0}(2400)$ 
is also important because  its properties influence 
the shape of the scalar form factor $f_0$
in semileptonic $D\to \pi$ decays~\cite{Flynn:2007ki,  Flynn:2000gd}, and indirectly it
has  some impact in the form factor $f_+$ that determines
$|V_{cd}|$~\cite{Burford:1995fc, Ball:2004ye, Dalgic:2006dt, 
Flynn:2007qd,Flynn:2007ii}. The bottom analogue is even more 
interesting because of the existing tension between the determinations of
$|V_{ub}|$ from inclusive and exclusive $\bar
B$ decays~\cite{Amhis:2014hma, Aoki:2016frl}, and the implications on 
 the unitarity triangle~\cite{Charles:2004jd, Bona:2006ah}
and on new physics limits~\cite{Bevan:2013kaa}.

We study here the heavy--light pseudoscalar meson $J^P=0^+$ scattering
in the strangeness-isospin $(S,I)=(0,1/2)$ sector, and present a
strong case for the existence of two poles in the $D_0^*(2400)$
energy region (and similarly in the bottom sector). The affirmative
evidence comes from a remarkably good agreement between our {\em
  parameter-free predicted} energy levels and the LQCD results
reported in Ref.~\cite{Moir:2016srx}. This two-pole structure was
previously claimed in Refs.~\cite{Kolomeitsev:2003ac,Guo:2006fu,Guo:2009ct}, and  finds
now a strong support. Its dynamical origin is
elucidated from the light-flavor SU(3) structure of the
interaction, and we find that the lower pole is the SU(3) partner of the
$D^\ast_{s0}(2317)$. Predictions for other $(S,I)$ sectors, including
bottom ones, will also be given.

\paragraph{Formalism.---} We consider the $S$-wave $D\pi$, $D\eta$, and $D_s
\bar{K}$ coupled-channel scattering.   A unitary $T$-matrix can be written as
\cite{Oller:1997ng,Oller:2000fj}:
\begin{equation}\label{eq:tmat}
T^{-1}(s) = V^{-1}(s) - \mathcal{G}(s)~,
\end{equation}
with $s \equiv E^2$, the center-of-mass (CM) energy squared. The
diagonal matrix $\mathcal{G}$ is constructed from the two-meson loop
function, $\mathcal{G}_{ii}(s) = G(s,m_i,M_i)$~\cite{Liu:2012zya},
where $m_i$ and $M_i$ are the masses of the light and
heavy pseudoscalar mesons in the channel $i$,  respectively. It carries the 
unitarity
cut and is regularized with a subtraction constant $a(\mu)$ (at a
scale $\mu = 1\ \text{GeV}$).  The matrix $V(s)$ contains the
interaction potentials, which are taken from the $\mathcal{O}(p^2)$
chiral Lagrangian of Ref.~\cite{Guo:2008gp}. They depend on six low
energy constants (LECs), $h_{0,\ldots,5}$. For the LECs and $a(\mu)$
we use the values and uncertainties obtained in
Ref.~\cite{Liu:2012zya} from a fit to LQCD results of the $S$-wave
charm--light pseudoscalar-meson scattering lengths in several $(S,I)$
sectors. Notice that the channel $(0,1/2)$ was not included in this
fit.

Above threshold, the phase shift [$\delta_i(s)$] and the inelasticity
[$\eta_i(s)$] of channel $i$ are related to the diagonal elements of
$T$, $i\, p_i\,T_{ii} =  4\pi\sqrt{s}\left(\eta_i\,
e^{2i\delta_i}-1\right)$,
with $p_i(s)$ the CM momentum. Bound, resonant, and virtual
states are associated to poles in different Riemann sheets (RS) of
the $T$-matrix. In our three-channel problem,  RS are denoted as
$(\xi_1\, \xi_2\, \xi_3)$, $\xi_i = 0,1$, and are defined
through analytical continuations:
\begin{equation}\label{eq:RS}
\mathcal{G}_{ii}(s) \to \mathcal{G}_{ii}(s) + i \frac{p_i(s)}{4\pi\sqrt{s}}\ 
\xi_i~.
\end{equation}
Thus,  $(000)$ is the physical RS. The coupling $g_i$ of a
pole to the channel $i$ is obtained from the residue ($g_i^2$) of  $T_{ii}$ .

In LQCD, interactions between quarks and gluons are considered in 
a cubic finite lattice with some boundary conditions. 
In a finite volume, the momentum running in loops  can take
only discrete values, $\vec{q} = \frac{2\pi}{L} \vec{n}$,
$\vec{n} \in \mathbb{Z}^3$, with $L^3$ the lattice volume. The
loop functions  are  replaced by
\cite{Doring:2011vk}:
\begin{align}
& \widetilde{\mathcal{G}}_{ii}(s,L) =
\mathcal{G}_{ii}(s)
+ \frac{1}{L^3} \sum_{\vec{n}}^{\left\lvert \vec{q\,}\right\rvert <
  \Lambda}\!\!\!
I_i(\vec{q}\,)- \int_0^\Lambda \!\! \frac{q^2
  \mathrm{d}q}{2\pi^2}\ I_i(\vec{q}\,)~,
\end{align}
with $\Lambda \to \infty$ and $I_i(\vec{q}\,)$ given in Eq.~(13) of
Ref.~\cite{MartinezTorres:2011pr}. The potentials $V(s)$
in Eq.~\eqref{eq:tmat} do not receive any finite volume
correction, and thus the $T$-matrix in a finite
lattice, $\widetilde{T}(s,L)$, reads:
\begin{equation}
\widetilde{T}^{-1}(s,L) = V^{-1}(s) - \widetilde{\mathcal{G}}(s,L)~.
\end{equation}
The energy levels are obtained as poles of $\widetilde{T}(s,L)$, and
they can be directly compared with those obtained in LQCD
simulations.

Here we compare with Ref.~\cite{Moir:2016srx}, where energy levels
relevant to the $D\pi$, $D\eta$, and $D_s \bar{K}$ channels at
different volumes are reported. We employ the meson masses of that
reference  in the evaluation of $V(s)$ and
$\mathcal{G}_{ii}(s)$. 
\begin{figure}[t!]\centering
\includegraphics[width=0.48\textwidth,keepaspectratio]{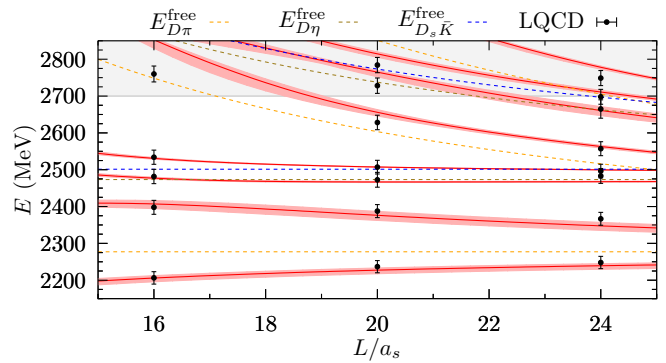}
\caption{Comparison of Ref.~\cite{Moir:2016srx}  $(0,1/2)$ energy
  levels (black dots) with our predictions
(red lines and bands). The bands represent the $1\sigma$
  uncertainties derived from the LECs fitted in
  Ref.~\cite{Liu:2012zya}, and $a_s= 0.12$ fm.  \label{fig:spectrum}}
\end{figure}

\paragraph{Results.---} The $(0,1/2)$ energy levels as a function of
$L$ are shown in Fig.~\ref{fig:spectrum}. The region above
$2.7\ \text{GeV}$ (shaded in Fig.~\ref{fig:spectrum}) is beyond the
range of applicability of our $\mathcal{O}(p^2)$ chiral unitary
formalism. Below that energy, the agreement of our computed energy
levels with those obtained in the LQCD simulation is excellent. This
is remarkable, since no fit to the LQCD energy levels is performed.

\begin{figure}[t]\centering
\includegraphics[width=0.48\textwidth,keepaspectratio]{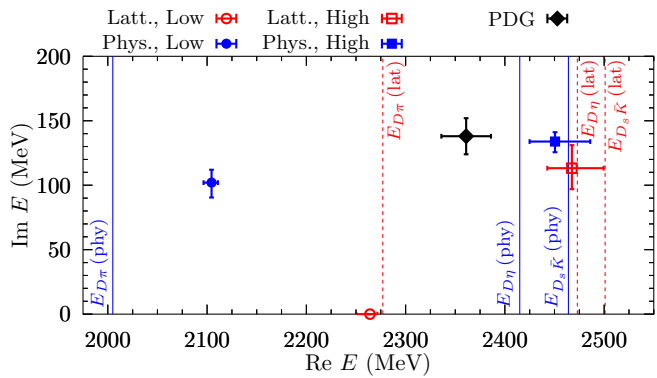}
\caption{Complex energy plane location of the two-pole-structure collected in
Table~\ref{tab:poles}. Empty red (filled blue) symbols stand for
the poles obtained when the LQCD~\cite{Moir:2016srx} (physical) masses
are used.  The black diamond represents the isospin average of the PDG values 
for $D^\ast_0(2400)^0$ and $D^\ast_0(2400)^+$ \cite{PDG2016}.%
\label{fig:poles}}
\end{figure}

The level below $D\pi$ threshold is interpreted in
Ref.~\cite{Moir:2016srx} as a bound state associated to the
$D^\ast_0(2400)$. For infinite volume and with the lattice meson
masses, our $T$-matrix also presents this pole.  The second
level, lying between the $D\pi$ and $D\eta$ thresholds, is very
shifted with respect to both of them, hinting at the presence of
another pole in infinite volume, that we find slightly below the
$D\eta$ threshold. Both poles are collected in the 
upper half of
Table~\ref{tab:poles} and represented
with empty red symbols in Fig.~\ref{fig:poles}.
\begin{table}[t]\centering
\begin{tabular}{ccccrrr}
Masses                      & $M$ (MeV)            & $\Gamma/2$ (MeV) & 
RS    & $\cou{D\pi}$ & $\cou{D\eta}$ & $\cou{D_s 
\bar{K}}$ \\
\hline
\multirow{2}{*}{lattice}    & $2264^{+\hphantom{1}8}_{-14}$ &   $0$             
       & (000) & $7.7^{+1.2}_{-1.1}$ & $0.3^{+0.5}_{-0.3}$ & 
$4.2^{+1.1}_{-1.0}$ \\
                            & $2468^{+32}_{-25}$            & $113^{+18}_{-16}$ 
       & (110) & $5.2^{+0.6}_{-0.4}$ & $6.7^{+0.6}_{-0.4}$ & 
$13.2^{+0.6}_{-0.5}$ \\ \hline
\multirow{2}{*}{physical}   & $2105^{+6}_{-8}$              & $102^{+10}_{-12}$ 
       & (100) & $9.4^{+0.2}_{-0.2}$ & $1.8^{+0.7}_{-0.7}$ & 
$4.4^{+0.5}_{-0.5}$ \\
                            & $2451^{+36}_{-26}$            & $134^{+7}_{-8}$   
       & (110) & $5.0^{+0.7}_{-0.4}$ & $6.3^{+0.8}_{-0.5}$ & 
$12.8^{+0.8}_{-0.6}$ \\ \hline
\end{tabular}
\caption{Position ($\sqrt{s}=M-i \Gamma/2$), couplings (in GeV) and RS of the two poles
  found in the $(0,1/2)$ sector using LQCD~\cite{Moir:2016srx}
  or physical masses.\label{tab:poles}}
\end{table}%

Next, we study the spectroscopic content of our amplitudes when the
physical masses are employed. The found poles are collected in
Table~\ref{tab:poles}, and shown in Fig.~\ref{fig:poles}. Comparing
the couplings, we see that the bound state below the $D\pi$ threshold
evolves into a resonance above it when physical
masses are used (notice that the threshold decreases from 2277 MeV to 2005 
MeV). Such an evolution is typically found for $S$-wave
poles ({\it e.g.}, for the $\sigma$ meson
\cite{Hanhart:2008mx,Albaladejo:2012te}). The second pole moves very
little and its couplings are rather independent of the meson
masses. For physical masses, it is a resonance located between the
$D\eta$ and the $D_s \bar{K}$ thresholds in the $(110)$ RS, 
continuously connected to the physical sheet. The mass of the
$D^\ast_0(2400)$ reported by the PDG \cite{PDG2016} lies between those
of the two poles found here, whereas the widths are similar
(Fig.~\ref{fig:poles}). We conclude the $D^\ast_0(2400)$ structure is
produced by two different states (poles), alongside with complicated
interferences with the thresholds.  This two-pole structure was
previously  reported in Refs.~\cite{Kolomeitsev:2003ac,Guo:2006fu,Guo:2009ct}, and
it receives here a robust support.

\begin{figure}[t]\centering
\includegraphics[width=0.48\textwidth,keepaspectratio]{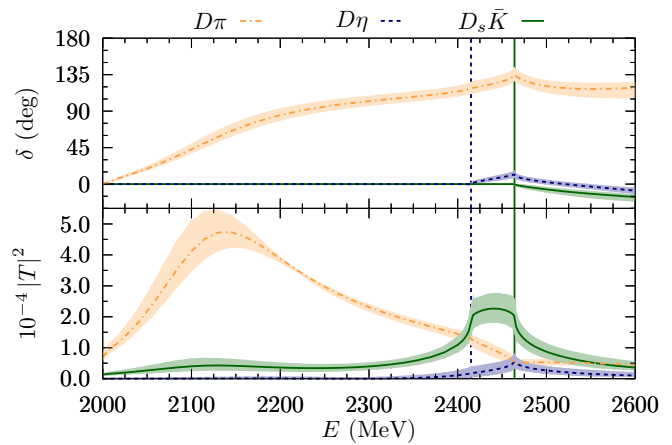}
\caption{Phase shift  and modulus squared of the diagonal
  $D\pi$, $D\eta$ and $D_s \bar{K}$  amplitudes, $T_{ii}(s)$,
in the $(0,1/2)$ sector.\label{fig:PhTsqInel}}
\end{figure}

Phase shifts and $\left\lvert T_{ii}(s) \right\rvert^2$ are shown in
Fig.~\ref{fig:PhTsqInel}. The lower pole causes a very mild effect in
the $D\eta$ and $D_s \bar{K}$ amplitudes. It couples mostly to $D\pi$
where a peak around $2.1\ \text{GeV}$ is clearly seen, while the phase
goes through $\pi/2$ at $\sqrt{s} \simeq 2.2\ \text{GeV}$. 
The higher pole manifests in a more subtle way. It produces a small
enhancement in the $D\pi$ amplitude, but the strongest effect is a
clear peak in the $D_s \bar{K}$ amplitude around
$2.45\ \text{GeV}$. However, the shape is quite
non-conventional. Despite the relatively large width, the amplitude
shows a narrow peak stretched between two cusps at the $D\eta$ and
$D_s \bar{K}$ thresholds. Such a behavior provides a possible test of the
two-pole structure. The BaBar and Belle data for $B\to D_s^-K\pi$ show an
enhancement in the $D_s^- K$ invariant mass
distribution~\cite{Aubert:2007xma,Wiechczynski:2009rg} (see also
Ref.~\cite{Antipin:2006ax}). This might confirm the features of the
second pole found here, although better statistics data are
required. 

\begin{figure}[t]
\includegraphics[width=0.48\textwidth]{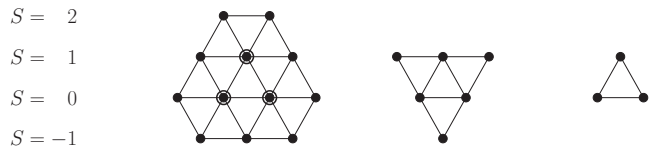}
\caption{Weight diagrams of the  $\mathbf{\overline{15}}$, $\mathbf{6}$ and 
$\mathbf{\overline{3}}$ irreps. \label{fig:irreps}}
\end{figure}

\paragraph{SU(3) study.---} To
gain further insights, we study the evolution of the two poles in the
light-flavor SU(3) limit, {\it i.e.}, when all light and heavy meson
masses take common values, $m_i = m$ and $M_i = M$,
respectively. Similar analyses were done in
Refs.~\cite{Kolomeitsev:2003ac,Gamermann:2006nm}. In this limit, the
heavy--light meson scattering decomposes into irreducible
representations (irreps) as $\mathbf{\overline{3}} \otimes \mathbf{8}
= \mathbf{\overline{15}} \oplus \mathbf{6} \oplus
\mathbf{\overline{3}}$ (Fig.~\ref{fig:irreps}), and the potential
matrix can be diagonalized, $V_d(s) = D^\dagger V(s) D = \text{diag}
\left( V_{\mathbf{\overline{15}}},
V_{\vphantom{\overline{\mathbf{6}}}\mathbf{6}},
V_{\mathbf{\overline{3}}} \right)$.  Since the three channels have a
common subtraction constant~\cite{Liu:2012zya}, the $T$-matrix is
diagonalizable, $T_d(s) = D^\dagger T(s) D = \text{diag}
(T_{\mathbf{\overline{15}}},
T_{\vphantom{\overline{\mathbf{6}}}\mathbf{6}},
T_{\mathbf{\overline{3}}})$, where $T^{-1}_A(s) =
V^{-1}_A(s)-G(s,m,M)$, $A \in \{ \mathbf{\overline{15}}, \mathbf{6},
\mathbf{\overline{3}} \}$. At chiral $\mathcal{O}(p)$, $V_d(s) =
f(s)\,\text{diag}(1,-1,-3)$, with $f(s)$ positive in the scattering
region, showing that the interaction in the $\mathbf{6}$ and
$\mathbf{\overline{3}}$ ($\mathbf{\overline{15}}$) irreps is
attractive (repulsive). The most attractive irrep, $\mathbf{\overline{3}}$, admits a $c\bar{q}$ ($q=u,d,s$) configuration. At
$\mathcal{O}(p^2)$, the potentials receive corrections, but these
qualitative features remain.

\begin{figure}[t]
\includegraphics[width=0.48\textwidth]{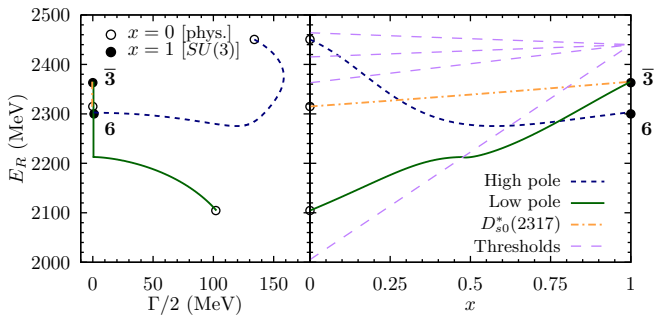}
\caption{Evolution from the physical to the flavor SU(3) symmetric
  cases of $(0,1/2)$ and $(1,0)$ poles $\sqrt{s} = E_R - i
  \Gamma/2$. The blue short dashed, green solid and the orange dashed
  dotted lines represent the higher and lower $D^\ast_0(2400)$ and the
  $D^\ast_{s0}(2317)$, respectively.  Left: path followed by the poles
  in the complex plane.  Right: evolution of $E_R$ with $x$. The
  purple long dashed lines stand for the
  $D\pi$, $D\bar{K}$, $D\eta$, and $D_s \bar{K}$ thresholds (from
  bottom to top).
  \label{fig:path}}
\end{figure}

We can connect the physical and SU(3) symmetric cases by
continuously varying the meson masses as:
\begin{align}
m_i & = m_i^\text{phy} + x(m-m_i^\text{phy})~, \label{eq:xevo1}
\end{align}
and analogously for $M_i$. Thus, $x=0$ ($x=1$) corresponds to the
physical (SU(3) symmetric) case. Numerically, we take $m =
0.49\ \text{GeV}$ and $M=1.95\ \text{GeV}$. The evolution of the poles
with $x$ is shown in Fig.~\ref{fig:path}. The lower $D^\ast_0(2400)$
pole found in the physical case (in the $(100)$ RS) connects with a
bound state of $T_\mathbf{\overline{3}}$ in the SU(3) limit, whereas
the higher pole (in the $(110)$ RS) connects with a virtual
($V_{\vphantom{\mathbf{\overline{6}}}\mathbf{6}}$ is not  
attractive enough to bind) state generated in
$T_{\vphantom{\overline{\mathbf{6}}}\mathbf{6}}$.\footnote{In the
  physical case, the RS are specified by $(\xi_1\,\xi_2\, \xi_3)$,
  with $\xi_i = 0$ or $1$ [Eq.~\eqref{eq:RS}]. In the SU(3) symmetric
  case, since all channels have the same threshold, there are only two
  RS, the physical [$(000)$] and the unphysical [$(111)$] sheets. To
  connect the lower pole in the physical case, located in the $(100)$
  RS, with the $T_\mathbf{\overline{3}}$ pole, in the $(000)$ RS, we
  vary the parameter $\xi_1 = 1-x$.  For the higher pole, one has to
  evolve $\xi_3 = x$ to connect $(110)$ (physical case) with $(111)$
  (SU(3) limit).}

In the $(S,I)=(1,0)$ sector
involving the $D{K}$ and $D_s \eta$ channels, and using the same
inputs and physical masses, we find a bound state 
at $2315^{+18}_{-28}\ \text{MeV}$~\cite{Liu:2012zya} which is
naturally identified with the $D^\ast_{s0}(2317)$. Its
evolution is also shown in Fig.~\ref{fig:path}, and we see it emerges
from the $T_\mathbf{\overline{3}}$ pole. Hence, the
$D^\ast_{s0}(2317)$ and the lower $D^\ast_0(2400)$ poles are flavour
SU(3) partners.

We now discuss other sectors in the physical case. The $(-1,0)$ involves only 
the
$D\bar{K}$ channel, and it is part of the $\mathbf{6}$ irrep, which is
weakly attractive. Indeed, we find a virtual pole, at
$2342^{+13}_{-41}\ \text{MeV}$, roughly $20\ \text{MeV}$ below
threshold, which has a sizable influence on the $D\bar{K}$
scattering length \cite{Liu:2012zya}.  The $(1,1)$ sector, involving the 
$D_s\pi$ and
$DK$ channels, has contributions from the $\mathbf{6}$ and
the repulsive $\mathbf{\overline{15}}$. Because of this, we do not find any pole
that can be associated to a physical state.

In the bottom sector, due to the heavy-flavor symmetry,\footnote{The
  LECs $h_i$ appearing in the Lagrangian of
  Refs.~\cite{Liu:2012zya,Guo:2008gp} depend on the heavy quark mass
  $m_Q$. By comparing with the Lagrangian of
  Ref.~\cite{Cleven:2010aw}, one deduces that the LECs scale as $h_i
  \sim m_Q$ for $i=0,\ldots,3$ and $h_i \sim 1/m_Q$ for
  $i=4,5$. Furthermore, we follow Ref.~\cite{Guo:2006fu} to translate
  the value of the subtraction constant from the charm to the bottom
  sector.}  we foresee a similar pattern.  In the $(0,1/2)$ sector
there is also a two-pole structure, located at
$(5537^{+9}_{-11},116^{+14}_{-15})\ \text{MeV}$ and
$(5840^{+12}_{-13},25^{+6}_{-5})\ \text{MeV}$. For $(S,I)=(1,0)$, we
find a state with a mass of $5724^{+17}_{-24}\ \text{MeV}$, bound by
about $50\ \text{MeV}$, as the $D^\ast_{s0}(2317)$ in the charm
sector.  All these pole positions are very similar to those found
already at $\mathcal{O}(p)$ \cite{Guo:2006fu}. In the $(-1,0)$ sector,
we find a virtual state located almost at threshold, which can also
appear as a bound state considering the $\mathcal{O}(p^2)$ parameter
uncertainties. As in the charm case, we do not find physical poles in
the $(1,1)$ sector that could be identified with the $X(5568)$,
recently reported by the D0 Collaboration \cite{D0:2016mwd}, but not
seen in other experiments ~\cite{Aaij:2016iev, CMS:2016fvl}. We
conclude the $X(5568)$ is not generated by the $\bar{B}_s
\pi-\bar{B}K$ rescattering~\cite{Albaladejo:2016eps} (see also
Refs.~\cite{Guo:2016nhb,Kang:2016zmv,Yang:2016sws}).

Finally, we remind that heavy quark spin symmetry relates the $0^+$
and $1^+$ sectors, and thus in the latter we find a similar pattern of
bound, resonant and virtual states~\cite{Kolomeitsev:2003ac,
  Hofmann:2003je,Guo:2006fu,Guo:2006rp,Cleven:2010aw}. We highlight
the predictions for the $\mathbf{\overline{3}}$ multiplet, where we
find $2436^{+16}_{-22}\ \text{MeV}$ and
$\left(2240^{+5}_{-6}-i\,93^{+9}_{-9}\right)$~MeV, for the
$D_{s1}(2460)$ and a new $D_1$ resonance, respectively. In the bottom
sector, we predict $5768^{+17}_{-23}\ \text{MeV}$ and
$\left(5581^{+9}_{-11}-i\,115^{+13}_{-15}\right)$~MeV for the $B_{s1}$
and the lowest $B_1$. The higher $D_1$ and $B_1$ poles stemming from
the $\mathbf{6}$ will presumably be affected by channels involving
$\rho$ mesons~\cite{Molina:2009eb, Soler:2015hna}. For $(-1,0)$, we
find, as in the $0^+$ case, an axial state located almost at threshold
in the bottom sector, while for charmed mesons 
the pole (virtual) moves deep in the complex plane.
 
\paragraph{Conclusions.---} We
have studied the $D\pi$, $D\eta$ and $D_s\bar{K}$ scattering in the
$J^P=0^+$ and $(S,I)=(0,1/2)$ sector. Although so far only one meson,
the $D_0^*(2400)$, has been reported in experiments~\cite{PDG2016}, we
present a strong support for the existence of two poles in the
$D_0^*(2400)$ mass region: the physical amplitudes, that contain two
poles, when put in a finite volume produce energy levels that
successfully describe the recent LQCD results in
Ref.~\cite{Moir:2016srx} without adjusting any parameter. The two
poles are located at
$\left(2105^{+6}_{-8}-i\,102^{+10}_{-12}\right)$~MeV and
$\left(2451^{+36}_{-26}-i\,134^{+7}_{-8}\right)$~MeV, with the largest
couplings to the $D\pi$ and $D_s\bar K$ channels, respectively.  A
group theoretical analysis shows that the lower pole and the
$D_{s0}^*(2317)$ complete the $\mathbf{\overline{3}}$ multiplet, 
being thus flavor SU(3) partners. We expect the two-pole structure to
produce distinctive features in $D\pi$, $D\eta$ and $D_s\bar K$
invariant mass spectra in high-energy reactions, such as $\bar B$
decays. In particular, despite of its large width, the higher pole
shows up as a narrow peak in the $D_s\bar K\to D_s\bar K$ amplitude
and should produce a sizable near-threshold  enhancement.
Note that clear $D_s\bar K$ threshold enhancements have been already
observed in $B$
decays~\cite{Aubert:2007xma,Wiechczynski:2009rg}. Future data from
better statistics experiments, such as LHCb and Belle-II, will shed
light into their origin.

A similar resonance pattern is also found in the bottom sector.
Besides the two-pole structure, we stress the possible existence of a
near-threshold bound or virtual state in the $\bar B \bar K$ (or $BK$)
channel, both for $0^+$ and $1^+$ sectors. These exotic states, with
quark content $bs\bar d \bar u$, will have a
large impact in the scattering length, and if bound they could only
decay through weak and/or electromagnetic interactions. 

The predicted phase shifts, both in the charm and bottom sectors, can
be used as inputs to the Omn\`es representation of the scalar form
factors describing heavy meson semileptonic decays~\cite{Flynn:2007ki,
  Flynn:2000gd}. Thus, the special two-pole structure discussed here
is also of interest to achieve an accurate determination of the
Cabibbo-Kobayahi-Maskawa matrix elements.

It is also worthwhile to notice the resemblance between the results
obtained here for the $D^*_0(2400)$ and the widely discussed two-pole
structure of the  $\Lambda(1405)$ linked to $\Sigma\pi$ and $N \bar K$~\cite{PDG2016,
  Oller:2000fj,Jido:2003cb,GarciaRecio:2003ks, Magas:2005vu}. The
existence of such a two-pole structure is rooted in both cases in
chiral dynamics, which on one hand determines the 
interaction strength, and on the other hand ensures the
lightness of  pions and kaons. The latter is important to separate
the two poles from higher hadronic channels. A two-pole structure 
driven by chiral dynamics is also found in Refs.~\cite{Roca:2005nm,
  Geng:2006yb,GarciaRecio:2010ki} for the $K_1(1270)$.

\begin{acknowledgments}
M.~A. and P.~F.-S. would like to thank the hospitality of CAS Key
Laboratory, where this work was initiated. M.~A. acknowledges
financial support from the ``Juan de la Cierva'' program
(27-13-463B-731) from the Spanish MINECO. This work is supported by the Spanish MINECO and European FEDER funds under the
contracts FIS2014-51948-C2-1-P, FIS2014-57026-REDT and SEV-2014-0398,
by Generalitat Valenciana under contract PROMETEOII/2014/0068, by DFG and NSFC through funds provided to the
Sino-German CRC 110 ``Symmetries and the Emergence of Structure in QCD'' (NSFC
Grant No.~11621131001), by the CAS Key Research Program of Frontier Sciences (Grant
No.~QYZDB-SSW-SYS013), and by the Thousand Talents Plan for Young Professionals.

\end{acknowledgments}

\bibliographystyle{plain}

\end{document}